%% file: sent_plb.tex
\documentclass[aps,prl,twocolumn,floatfix,superscriptaddress,nofootinbib]{revtex4}
\usepackage{graphicx,amsmath}

\usepackage{graphicx}
\usepackage[figuresright]{rotating}

\begin{document}

\title{Quark fragmentation to $\pi^{\pm}$, $\pi^{0}$, $K^{\pm}$, $p$ and $\bar{p}$ in the nuclear environment}



\include{autori_qm}

%

\begin{abstract}
 The influence of the nuclear medium on lepto-production of  hadrons 
 was studied in the HERMES experiment at DESY in semi-inclusive deep-inelastic scattering of 27.6 GeV positrons
off deuterium, nitrogen and krypton targets.
 The differential multiplicity for krypton relative to that of deuterium has been measured 
for the first time for various identified hadrons ($\pi^+$, $\pi^-$, $\pi^0$, $K^+$,  $K^-$, 
$p$ and $\bar{p}$)
as a function of 
 the virtual photon energy 
$\nu$, the fraction $z$ of this energy transferred to the hadron,
and the hadron transverse momentum squared $p_t^2$.
The multiplicity ratio is strongly reduced in the nuclear medium at low $\nu$ and high $z$, 
with significant differences among the various hadrons.
 The distribution of the  hadron transverse momentum 
is broadened towards high $p_t^2$ in the nuclear medium,
in a manner resembling
 the Cronin effect previously observed in collisions of heavy ions and  protons with nuclei.
\end{abstract}
\maketitle




\vspace{.2cm}
The phenomenon of confinement in QCD imposes itself dynamically
in the process 
 of hadronization, i.e.
the mechanism  by which  final-state hadrons are
formed from a  quark that has been struck hard.
This process,  also known as the fragmentation of quarks 
into hadrons, can be described by  fragmentation functions 
$D_f^h(z)$,  
 denoting the probability that a quark of flavor $f$ produces a hadron of type $h$
carrying a fraction $z$  
 of the energy of the struck quark in the target rest frame.
In the nuclear medium additional soft processes may occur  
before the final-state hadron is completely formed.
The nuclear environment may thereby influence the hadronization process, 
e.g. cause a change in the quark fragmentation functions, in analogy to the EMC finding of 
a medium modification of the quark distribution functions~\cite{AUB83}.

The understanding of quark propagation  in the nuclear medium
is crucial for the interpretation  of  ultra-relativistic heavy ion collisions,
as well as high energy proton-nucleus  and  lepton-nucleus interactions~\cite{BAI00}.  
Quark propagation in the nuclear environment
 involves  processes like multiple interactions with the surrounding medium and induced gluon radiation, 
resulting in energy loss of the quark.
If the final hadron is formed inside the nucleus, the hadron can interact via the relevant hadronic interaction cross section, 
causing a further reduction of the hadron yield. Therefore,
quark and hadron propagation in nuclei are expected to result in a modification i.e. a ``softening'' of the leading-hadron 
spectra~\cite{NI79} compared to that from a
free nucleon.
By studying the properties of the leading-hadrons emerging from nuclei,
 information on the characteristic time-distance scales of 
hadronization can be derived.

The hadronization process in the nuclear medium is traditionally  described in the framework 
of phenomenological string models~\cite{BI80,GY90,CZ92,AK02} and
final state interactions of the produced
hadrons with the surrounding medium~\cite{FAL03}.
Alternatively, in-medium modifications of the quark fragmentation functions 
have  been proposed, either expressed in terms of  
their nuclear rescaling~\cite{DE85,ACC02},
or parton energy loss~\cite{arleo} and  higher-twist contributions to the fragmentation functions~\cite{GUO04},
 or in terms of a gluon-bremsstrahlung model for leading hadron attenuation~\cite{boris}.
The models of Refs.~\cite{ACC02,boris} also incorporate hadronic final state interactions.
These recent  QCD-inspired models provide a theoretical description of the hadronization process in deep-inelastic scattering, 
relativistic heavy-ion collisions \cite{GUO04} and Drell-Yan reactions on nuclear targets~\cite{arleo,DY}.
Moreover, some of these models contain a so-far untested QCD prediction, i.e. that the induced radiative energy loss 
of a quark traversing a length $L$ of hot or cold matter is 
proportional to $L^2$~\cite{BA97}
due to the coherence of the gluon radiation process~\cite{LPM}.

Semi-inclusive deep-inelastic lepton-nucleus collisions  
are most suitable to obtain quantitative information on the
hadronization process.  In contrast to hadron-nucleus and nucleus-nucleus scattering, in 
deep-inelastic scattering
 no deconvolution of the parton distributions of the projectile and target particles is needed, 
so that hadron distributions and multiplicities 
from various nuclei can be
directly related to nuclear effects in quark propagation and hadronization.

\begin{figure}[htb]
\includegraphics[width=\columnwidth]{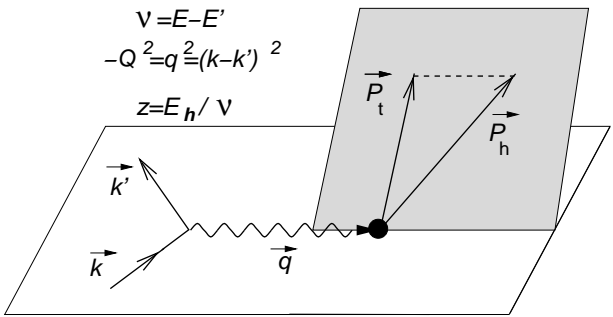}
\caption{Kinematic planes for hadron production in semi-inclusive deep-inelastic scattering
and definitions of the relevant lepton and hadron variables. The quantities $k$ ($k'$) and $E$ ($E'$)
 are  the 4-momentum and the energy of 
the incident (scattered) positron, and $p_h$ and $E_h$ are the 4-momentum and the energy of the produced hadron.}
\label{fig:fig00}
\nopagebreak
\end{figure}

The experimental results for semi-inclusive deep-inelastic scattering
on nuclei are 
usually presented in terms of the hadron
multiplicity ratio $R_M^{h}$, which 
represents the ratio of the number of hadrons of type $h$
produced per deep-inelastic scattering event on
a nuclear target of mass A to that from a deuterium target (D). 
 The ratio $R_M^{h}$  depends  on the leptonic variables $\nu$ and $Q^2$, 
where $\nu$ and $-Q^2$ are the energy in the target rest frame and the squared four-momentum 
of the virtual photon respectively, and on the hadronic variables 
$z$ and $p_t^2$, where  
$p_t$ is the hadron momentum component transverse to the virtual photon direction. 
Fig.~\ref{fig:fig00} illustrates the definition of the relevant lepton and hadron kinematic variables
 for this analysis.
The multiplicity ratio is defined as:
\vspace{-.0cm}
\begin{equation}
 R_M^{h}(z,\nu, p_t^2, Q^2) = {\frac {\left. \frac{N_h(z,\nu, p_t^2, Q^2)}{N_e(\nu, Q^2)}\right|_A }
			{\left. \frac{N_h(z,\nu, p_t^2, Q^2)}{N_e(\nu, Q^2)}\right|_D }},
\label{eq:att}
\end{equation}
\noindent
where  $N_h$ is the yield of semi-inclusive hadrons in a 
given ($z,\nu$,$p_t^2$,$Q^2$)-bin, and 
$N_e$ the yield of inclusive deep-inelastic scattering leptons in the same ($\nu$,$Q^2$)-bin.
The ratio  $R_M^{h}$ is usually evaluated as a function of $\nu$ and $z$ only,
while integrating over all other kinematic variables,
as existing data for $R_M^{h}$ show  a weak dependence on either  $Q^2$ or $p_t^2$~\cite{hunen,EMC}.

In the past, semi-inclusive  leptoproduction of 
undifferentiated hadrons from nuclei was studied 
at SLAC with electrons~\cite{osborne},
and at CERN and FNAL with high-energy muons by EMC~\cite{EMC} and
E665~\cite{Xe}. 
Recently, HERMES  reported more precise data~\cite{AIR01} on the 
production of charged hadrons, as well as identified $\pi^+$ and  $\pi^-$ mesons, in deep-inelastic positron scattering on
nitrogen relative to deuterium.
A significant difference was found  between the multiplicity ratio of positive and negative hadrons,
while the multiplicity ratio for identified pions was found to be the same for both charges.
In order to clarify this issue, additional 
measurements of $R_M^h$ 
with identification  of other hadron species 
have been performed at HERMES.

In this paper we present  results on the hadron multiplicities on krypton relative to deuterium,
  providing the first measurements of the multiplicity ratio for
identified pions, kaons, protons and antiprotons.
Additionally the nitrogen data for charged hadrons and identified pions are reevaluated, now covering  a wider kinematic range 
than in  Ref.~\cite{AIR01}.
The measurements described were performed with the HERMES spectrometer~\cite{hermesdetector} using the 27.6 GeV positron beam
stored in the HERA ring at DESY.
The spectrometer consists of two identical halves located above and below the positron  beam pipe.
Both the scattered positrons and the  produced hadrons were detected simultaneously within an angular acceptance  of  
$\pm$ 170 mrad horizontally, and $\pm$ 
(40 -- 140)  mrad vertically. 

The data were collected  using 
krypton and deuterium
 gas targets internal to the  positron storage ring.
Either polarised deuterium or unpolarised high density krypton gas was injected into a
   40 cm long tubular open-ended storage cell. 
Target areal
densities up to  1.4 $\times$ 10$^{16}$ nucleons/cm$^2$ were obtained for krypton.
During these high-density runs HERA operated in a dedicated mode for the HERMES experiment.
This made it possible to accumulate the krypton statistics within a few days in 1999.
The deuterium data were collected  over a period of one year (1999) with a lower-density  polarised target.
The yields from deuterium were  averaged over the two spin orientations.

The positron trigger was formed by a coincidence between signals from three scintillator hodoscope planes, and a lead-glass calorimeter
where a minimum energy deposit of 3.5 GeV (1.4 GeV) for unpolarised (polarised) target runs  was required.
The identification of the  scattered positrons 
was accomplished using 
 a transition-radiation detector, a scintillator
preshower counter, and an  electromagnetic calorimeter.

The identification of charged pions, kaons, protons and antiprotons 
 was accomplished using the information from the  RICH detector~\cite{RICH}, 
which replaced a threshold  \v Cerenkov counter used in the previously reported measurements on $^{14}$N~\cite{AIR01}.
This detector uses two radiators,
 a 5 cm thick wall of silica areogel tiles followed by
a large volume of    C$_4$F$_{10}$ gas,  to provide  separation of pions, kaons, and protons over most of the kinematic
 acceptance of the spectrometer.
The pions and kaons identified by the RICH detector are analysed in the momentum region between 2.5 GeV and 15 GeV,
while for the identified protons and antiprotons the momentum region is restricted to the range between 4 GeV and 15 GeV 
 in order to reduce  possible  contaminations from misidentified hadrons.
The identification efficiencies and contaminations for pions, kaons, protons and antiprotons have been  determined in
a Monte Carlo simulation as a function of the hadron momentum and multiplicity
in the relevant detector half.
These RICH performance parameters
were verified in a limited kinematical domain using known particle
species from identified resonance decays.
They were used in a matrix method to unfold the true hadron
distributions from the measured ones.

The electromagnetic calorimeter~\cite{calo} provided neutral pion identification by the
detection of  two neutral clusters originating  from the two decay photons.
Each of the two clusters was required to have an energy
$E_{\gamma}\ge0.8$ GeV.  
The background was evaluated in each kinematic bin by fitting the two-photon 
invariant mass spectrum with a Gaussian plus a polynomial that
fits  the shape of the background due to uncorrelated
photons.  
 The number of  detected $\pi^0$ mesons was obtained by integrating the peak, corrected for background, 
over the $\pm$2$\sigma$ range with respect to the centroid of the Gaussian. 
The identified neutral pions were analysed in the same momentum range
as the charged pions, i.e. 
 between 2.5 GeV and 15 GeV.

Scattered positrons  were selected by imposing the constraints 
$Q^2>1$~GeV$^2$, $W=\sqrt{2M\nu+M^2-Q^2} >2$~GeV
for the invariant 
mass of the photon-nucleon system where $M$ is the nucleon mass,    
and $y=\nu/E<$~0.85 for the energy 
fraction of the
virtual photon.
The requirements on $W$ and $y$ are applied to exclude nucleon resonances 
and to limit the magnitude of the radiative corrections, respectively.
In the previously reported data on  $^{14}$N~\cite{AIR01} an additional constraint
$x=\frac{Q^2}{2M\nu} >0.06$ for the Bjorken scaling variable
was applied, in order  to exclude the 
kinematic region  where an anomalous ratio of the longitudinal to 
transverse cross sections for deep-inelastic scattering from $^{14}$N appeared.
Since that report this anomaly was found to be due to a peculiar local instrumental inefficiency,
for which corrections  have now been evaluated~\cite{erratum}. By applying these corrections 
the $x$-range was extended down to $x$=0.02 both for the new data and  for the  previously 
published  $^{14}$N data.

Most of the hadrons from target fragmentation were suppressed by the requirement $z>$ 0.2.
The lower constraint on the  hadron momentum 
implies 
that the $z-$acceptance is restricted to rather large values
as $\nu$ decreases. Hence, to ensure that the $\nu$-dependence of the multiplicity ratio
does not correspond to a strong variation of the mean $z$-value, 
 the present data were confined to $\nu>$ 7 GeV.

Under the kinematic constraints described above, the number of
selected deep-inelastic scattering  events is 8.2 (7.3) $\times 10^5$ for 
krypton (deuterium).
The number of charged pions is 
11.3 (13.5) $\times 10^4$, while the number of neutral pions is 1.9 (2.3) $\times 10^4$.
The number of kaons is 
2.1 (2.3) $\times 10^4$,
and the number of protons and antiprotons 
is 1.1 (1.1) $\times 10^4$.

The  data have been corrected for radiative processes
involving nuclear elastic, quasi-elastic
 and inelastic scattering,
using the codes of Refs.~\cite{Tera,Akush}. 
The code of Ref.~\cite{Akush}
was modified to include the measured semi-inclusive deep-inelastic scattering cross sections.
The size of 
 the radiative corrections applied   to $R_M^h$ was found to be negligible
in most of the kinematic range, with a maximum of
about 7\% at the highest
value of $\nu$, as most of  the radiative contributions 
cancel in the multiplicity ratio~\cite{GAR02}.

The charged pion sample is contaminated by pions 
originating from the decay of heavier mesons.
The main effect on the pion multiplicities is due to the
 decay of exclusively produced  $\rho^0$ vector mesons,
which may affect the multiplicities by an amount ranging
from about 1\% at low $z$ up to 30\% (45\%) at high $z$ for
positive (negative) pions,
 as has been estimated from a  Monte Carlo simulation. 
 The effect on  the super-ratio $R_M^h$ is smaller, but does not cancel as the  $\rho^0$  
vector meson are also attenuated in the nuclear medium.
By taking into account the measured  $\rho^0$  
nuclear transparency~\cite{CT02},
the remaining effect 
on  $R_M^h$  has been  estimated and  included
 in the systematic uncertainty.
No correction  was applied to  $R_M^h$.
Pions resulting from the decay of $\rho^0$ mesons formed in the fragmentation process
are included in $R_M^h$.

The systematic uncertainty is reduced due to the fact that super-ratios
of seminclusive and inclusive yields are measured.
The contributions
to the systematic uncertainty of $R_M^h$ arise from radiative corrections 
($<$ 2\%),
 hadron identification (1.5\% for neutral pions,  0.5\%  for kaons,    1\%  for protons and  2\%  for antiprotons), overall efficiency ($<$ 2\%),
and  $\rho^0$-meson production for positive  (0.3\% - 4\%) and negative (0.3\% - 7\%) pions~\cite{GAR02}.

The geometric acceptance
for semi-inclusive hadron production has been verified to be the same for both 
 the krypton and deuterium 
targets
by studying the multiplicity ratio as a function of the
hadron polar angle.
This ratio was found to be constant  within the 
experimental precision. 

\begin{figure}[!tb]
\includegraphics[width=1.\hsize]{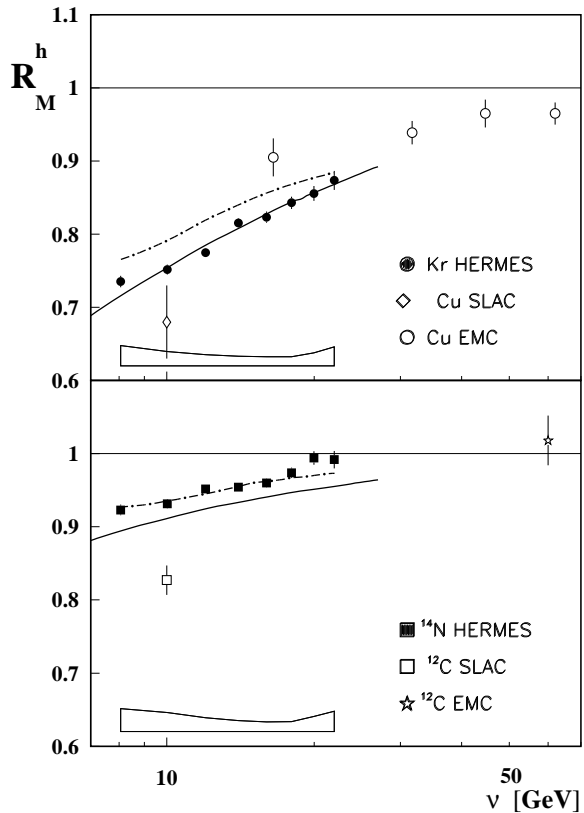}
\caption{Charged hadron multiplicity ratio $R_M^{h}$
as a function of $\nu$ for  $z>$ 0.2. In the upper panel  HERMES data on Kr are compared to
  SLAC~\cite{osborne} and CERN~\cite{EMC} 
data on Cu. 
In the lower panel the HERMES data on $^{14}$N are compared
with   CERN and SLAC data on $^{12}$C. The error bars
represent the statistical uncertainties, 
and the systematic  uncertainty is shown as the band.
The solid curves are calculations from Ref.~\cite{ACC02}  and the dot-dashed curves are calculations from Ref.~\cite{GUO04}.}
\label{fig:fig1}
\nopagebreak
\end{figure}

The multiplicity ratio has been determined  as a function of either $z$, $\nu$ or $p_t^2$, while integrating over all
 other kinematic variables.
In Fig.~\ref{fig:fig1}  the multiplicity ratios for
all charged hadrons with $z>$ 0.2
are presented as a function of $\nu$
 together with
data of previous experiments on nuclei of similar size.
 In the top panel 
the  HERMES data on  Kr  are compared 
with
 the SLAC~\cite{osborne} and CERN~\cite{EMC} 
data for Cu.
In the lower panel the reevaluated HERMES data on $^{14}$N are
displayed together with data  on $^{12}$C~\cite{osborne,EMC}. Due to the extension down to $x$=0.02,
the $^{14}$N data shown in  Fig.~\ref{fig:fig1} have a  higher statistical accuracy than the data reported in Ref.~\cite{AIR01}.
The HERMES data for $R_M^{h}$ are observed to increase with
increasing $\nu$, roughly approaching the EMC results at higher values of $\nu$.
The discrepancy with the SLAC data is partially due to the fact that  semi-inclusive
cross section ratios were measured at SLAC instead of  multiplicity ratios.
Thus no corrections in the SLAC data
were made for the target-mass dependence of the inclusive deep-inelastic scattering cross section,
as discussed in Ref.~\cite{AIR01}.

\begin{figure}[!tb]
\includegraphics[width=1.\hsize]{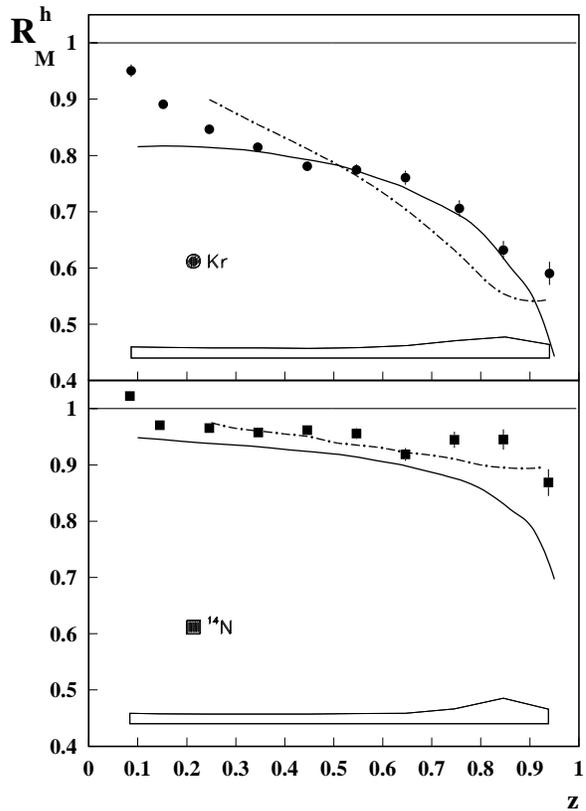}
\caption{Charged hadron multiplicities $R_M^{h}$ as a function of $z$ for $\nu >$ 7 GeV.
The data are compared to the same calculations as in Fig.~\ref{fig:fig1}.
 The error bars
represent the statistical uncertainties and the systematic uncertainty is shown as the band.}
\label{fig:fig2}
\nopagebreak
\end{figure}

A stronger attenuation 
is observed for Kr than for
$^{14}$N, the average ratios 
and the total experimental uncertainties 
being $R_M^{h}$=0.802 $\pm$ 0.021 and  $R_M^{h}$=0.954 $\pm$ 0.023, respectively.
 Fig.~\ref{fig:fig2} shows the dependence on $z$ of the same multiplicity ratios for $\nu >$ 7 GeV.
This figure includes the region  $z<0.2$, which contains  contributions  from 
both  target fragmentation hadrons and  leading hadrons decelerated 
in  nuclear re-scattering. 
Qualitatively, the dependences on  $\nu$ and $z$ of the Kr data resemble those of the $^{14}$N data, but
 the features are more pronounced.

The measured $\nu$- and $z$-dependences  for both 
krypton and nitrogen
are compared to  several model calculations   shown in Figs.~\ref{fig:fig1} and~\ref{fig:fig2} as  solid  curves~\cite{ACC02},  and
 dot-dashed curves~\cite{GUO04}.
In Ref.~\cite{ACC02} the nuclear modification of hadron production in deep-inelastic scattering  is described as a rescaling of the 
quark fragmentation functions, supplemented by nuclear absorption. 
In this model 
 the nuclear absorption contribution is dominated by
 the string interaction,
while the subsequent interaction of the fully formed hadron contributes only a few percent to $R_M^h$  for  krypton.
 The  calculation  
  overestimates  the 
$^{14}$N attenuation, but gives a fairly good account of
 both  the $\nu$ and $z$ dependences, for $z>$ 0.2, of the Kr data.
In Ref.~\cite{GUO04} 
  nuclear modification of the quark fragmentation process in  deep-inelastic scattering 
has been evaluated taking into account multiple parton scattering and induced energy
 loss in the medium.
The only free parameter of this model was tuned to reproduce the $^{14}$N data, 
and the derived  energy loss 
has been used to determine the initial gluon density in  the Au+Au collision data collected by the PHENIX
 collaboration at RHIC~\cite{PH01}.
The model roughly reproduces 
the changes in $R_M^h$ when going from nitrogen to krypton.
  
\begin{figure*}[p]  \center
\hspace{-0.4cm}
  {\includegraphics[width=8cm, height=19cm]{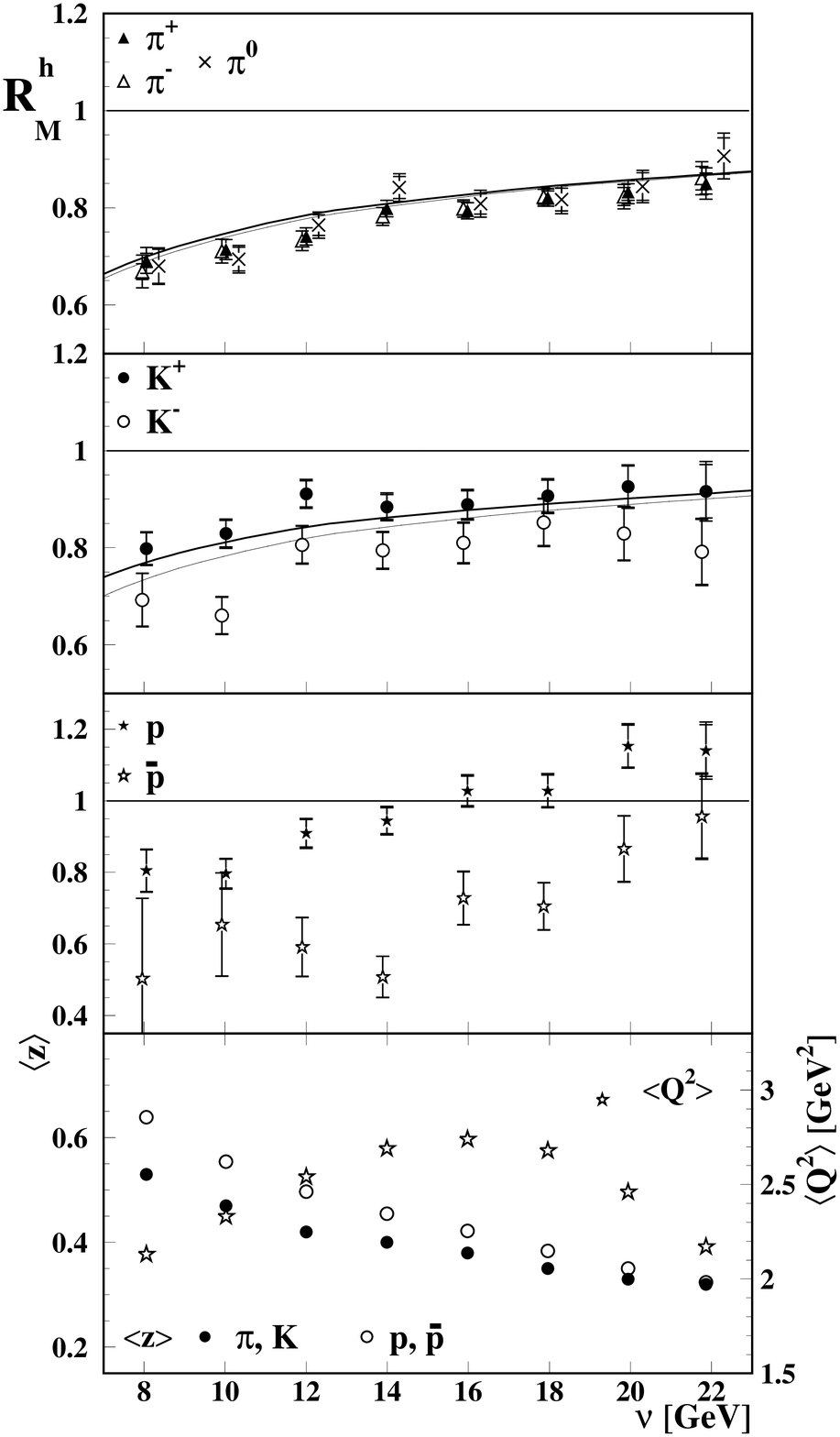}}
 {\includegraphics[width=8cm, height=19cm]{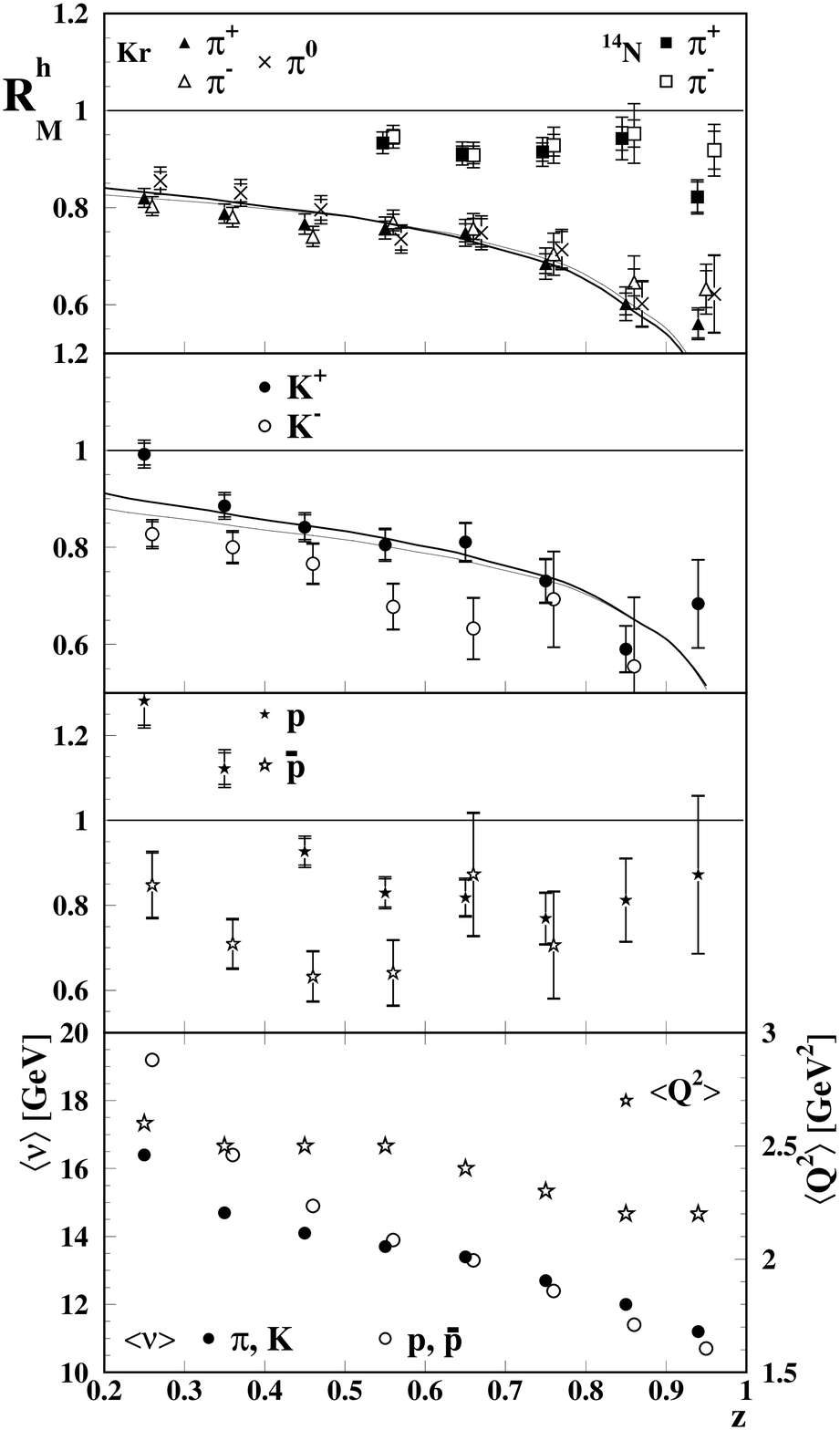}}
\caption{Multiplicity ratios for identified pions, kaons, protons and antiprotons from a Kr target as a function of $\nu$
 for  $z> $ 0.2 (left), and as a function of $z$ for  $\nu> $ 7 GeV (right). 
In the upper right panel the multiplicity ratio for identified pions  from a $^{14}$N target are also shown.
The closed (open) symbols represent the positive (negative)
	charge states, and the crosses represent $\pi^0$ mesons.  
In the bottom panels the average  $z$ and $\nu$ values are displayed: pions and kaons (protons and antiprotons)
are shown as closed (open) circles;
the average $Q^2$ values are indicated by the open stars referring to the right-hand scales.
The inner (outer) error bars represent the statistical (total) uncertainties.
The thick (thin) solid curves represent the calculations of Ref.~\cite{ACC02} for positive (negative) charge states.
Multiplicity ratios for negative kaons and antiprotons at the highest $z$-bins are not displayed due to their poor
statistical significance. }
\label{fig:fig3}
 \end{figure*}

For the first time the $\nu$-dependences of the multiplicity ratio were studied  for identified neutral and charged  pions, 
kaons,
 protons and antiprotons,  
as shown in the left part of Fig.~\ref{fig:fig3}.
The corresponding $z$-dependences of  $R_M^h$  
with $\nu>$ 7 GeV are shown in the right part of Fig.~\ref{fig:fig3}. 
In the bottom panels 
the average values
  for $Q^2$ and  $z$ or $\nu$ are displayed
for all the presented data. 
The results for both charge states of the pion and the kaon are compared to the calculations of Ref.~\cite{ACC02}.
A good description for the  $\pi^{\pm}$ and $K^+$ data is observed, while
 the  attenuation 
is under-predicted for the $K^-$ data. 
No model predictions are available for protons and antiprotons.
Average  $R_M^h$ values are obtained 
by integrating yields over $\nu$  and $z$. 
The results presented in   Fig.~\ref{fig:fig3} and the average  $R_M^h$  values reported
in 
Table~\ref{Table1} for  $z >$ 0.2, 
 show that the multiplicity ratios for
positive and negative pions are similar, in agreement with
what  was already found on  $^{14}$N~\cite{AIR01}. In addition, the multiplicity ratio
for neutral pions is
found to be  consistent, within the total  experimental uncertainties, with that for  charged pions
as well as for negative kaons.
However,  
 $R_M^{h}$ for positive kaons is significantly larger.
An even larger  difference is observed  between
protons and their antiparticles compared to the meson case.
These differences in  $R_M^{h}$ of positive and negative kaons, as well as those between protons and antiprotons,
 are  still present at higher values of $z$. This is shown in the last column of Table~\ref{Table1},
where the average $R_M^{h}$   values are  reported for $z>$ 0.5, i.e. when emphasising leading hadrons.
In addition the  $z >$ 0.5 range is most suitable to compare  $R_M^{h}$ of mesons and baryons 
as this comparison is performed at the same average $\nu$ as shown in
the bottom right panel of  Fig.~\ref{fig:fig3}.
\vspace{-.0cm}
\begin{table}[htb]
\begin{center}
\begin{tabular}{|c|c|c|} \hline 
 h-type & $<R_M^h>$ &  $<R_M^h>$ \\ 
        & $z>0.2$  & $z>0.5$ \\ \hline
$\pi^{+}$ & $0.775\pm 0.019$ & $0.712 \pm 0.023$ \\
$\pi^{-}$ & $0.770\pm 0.021$ & $0.731 \pm 0.031$ \\
$\pi^{0}$ & $0.807\pm 0.022$ & $0.728 \pm 0.024$ \\
$K^{+}$ & $0.880\pm 0.019$ & $0.766 \pm 0.024$ \\
$K^{-}$ & $0.783\pm 0.021$ & $0.668 \pm 0.036$ \\
$p$ & $0.977\pm 0.027$ & $0.816 \pm 0.029$ \\
$\bar{p}$ & $0.717\pm 0.038$ & $0.705 \pm 0.067$ \\
\hline
\end{tabular}
\caption{Multiplicity ratio values for krypton and deuterium  yields 
integrated over $z$ and $\nu >$ 7 GeV.
Total experimental uncertainties are quoted.}
\label{Table1}
\nopagebreak
\end{center}
\end{table}

 It has been 
  suggested ~\cite{GUO04} that the observed
 differences in $R_M^{h}$ 
can  be attributed to  the mixing of quark and gluon fragmentation functions.
This mixing gives a different modification of the quark and antiquark fragmentation
functions in nuclei,
 thus leading to a more significant difference between the multiplicity ratio of protons and antiprotons
than between those of mesons.
The observed differences in the multiplicity ratios  can also be interpreted in terms of different
formation times of baryons and mesons~\cite{KOP85},
or in terms of different hadron-nucleon interaction cross sections~\cite{PDG}.
While this cross section is 
 similar for positive and negative pions, 
it is larger for negative  kaons as compared to positive kaons,
and even larger for antiprotons
than  protons, in qualitative agreement with the trend shown by the data. 

The data for identified charged  pions   produced on  Kr and $^{14}$N 
for $z >$ 0.5 were used to estimate  the mass-number 
dependence of the 
nuclear attenuation.
 The nitrogen data, measured in a more restricted momentum range between 4--13.5 GeV due to use of the  \v Cerenkov  detector,
are  shown in the upper right panel of 
 Fig.~\ref{fig:fig3}. 
By using the same constraints on the krypton data 
and assuming a simple $A^{\alpha}$-dependence of the nuclear attenuation  $1-R_M^h$,
the experimental data are found to be closer to
 the $A^{\frac{2}{3}}$-dependence~\cite{GAR02} predicted  in Ref.~\cite{GUO04}, than 
 the   $A^{\frac{1}{3}}$-dependence  that follows from models 
based on nuclear absorption effects only.
Data on more nuclei are needed to enable systematic studies of the 
$A$-dependence of the nuclear attenuation effects.

\begin{figure}[!tb]
\includegraphics[width=1.\columnwidth]{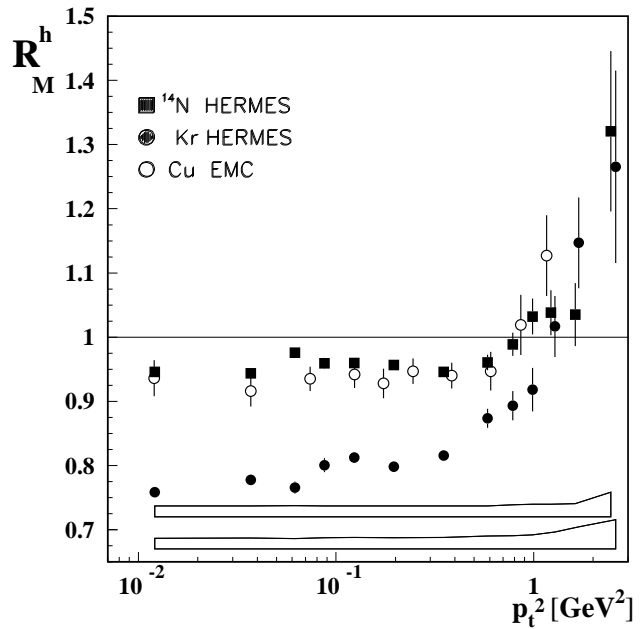}
\caption{Multiplicity ratio for charged hadrons  versus $p_t^2$ for  $\nu > $7 GeV and $z > $0.2. The HERMES data on Kr 
and  $^{14}$N are compared to the EMC~\cite{EMC} data for Cu in the range 10 $<\nu<$ 80 GeV.
The error bars
represent the statistical uncertainties. The systematic uncertainty for Kr ($^{14}$N) is shown as the lower (upper) band.} 
\label{fig:fig5}
\end{figure}

The $p_t$ distribution of the observed hadrons is expected to be broadened on a nuclear target  compared to a proton 
target
due to   multiple  scattering of the propagating quark and hadron.
This effect is known as the Cronin effect~\cite{CR75} and has previously been observed in heavy-ion and hadron-nucleus 
induced reactions.
A nuclear enhancement at high $p_t^2$ is also observed in the HERMES data shown in 
 Fig.~\ref{fig:fig5} both for $^{14}$N and Kr nuclei.
In this plot  the  EMC~\cite{EMC} data on Cu which 
cover a different $\nu$-range, 10 $<\nu<$ 80 GeV, are also displayed.
The data for  $p_t^2 <0.7$ GeV$^2$ show the attenuation previously discussed,
while the data for 
$p_t^2 \ge 0.7$ GeV$^2$
reflect the $p_t$-broadening 
ascribed to multiple scattering effects.
This effect  is similar to  the one reported for proton-nucleus and nucleus-nucleus collisions~\cite{CERN}
 but  is smaller in magnitude.
 The enhancement 
 is also predicted to occur at  a $p_t$-scale of about 1-2 GeV~\cite{WA01,KOP02},
 in agreement with  the semi-inclusive  deep-inelastic scattering data shown in Fig.~\ref{fig:fig5}. 
The HERMES data 
may help  to interpret the 
  new relativistic heavy-ion results from
 SPS~\cite{HI01} and RHIC~\cite{PH01}, which show
a weaker $p_t$ enhancement 
 than expected from the original Cronin effect.

In summary, the  multiplicities  of charged hadrons and of identified pions, kaons, protons and antiprotons 
on  krypton relative to deuterium were measured for the first time.
The data show that the multiplicity ratio $R_M^h$ is reduced at low $\nu$ and high $z$.
Different multiplicity ratios were observed for various hadrons. 
 In contrast to the similarity between positive and negative pions, a significant difference in $R_M^h$ is found between positive and negative 
kaons and a  larger difference between protons and antiprotons.
The different results for various hadrons may reflect differences in the modification of quark and antiquark
fragmentation functions~\cite{GUO04} and/or in the hadron nucleon interaction cross sections.

The hadron multiplicity is observed to  be enhanced at high $p_t^2$ in the nuclear medium,
showing evidence of  the Cronin effect in 
deep-inelastic scattering process.
This effect is similar to the one observed in hadron nucleus scattering, with a rise of $R_M^h$
to values above unity for $p_t^2 \geq$ 1 GeV$^2$. 

Additional measurements of differential hadron multiplicities on  both light and heavy nuclei with pion,
kaon and proton identification are underway at HERMES. Such measurements will also help to clarify
the issues raised by the present data concerning the mass-number dependence of the hadronization process
and of the Cronin effect for various identified hadrons.

\vspace{1.cm}
We thank A. Accardi, F. Arleo, W. Cassing, T. Falter,
 B.Z. Kopeliovich, U. Mosel,  J. Nemchik,  H.J. Pirner and X.N. Wang for many interesting discussions
on this subject.
We gratefully acknowledge the DESY management for its support, the staffs at
DESY,  and the collabo\-rating institutions for their significant effort.
This work was
supported by the FWO-Flanders, Belgium; the Natural Sciences and Engineering
Research Council of Canada; the ESOP, INTAS and TMR network contributions from
the European Union; the German Bundesministerium f\"ur Bildung und
Forschung; the Italian Istituto Nazionale di Fisica Nucleare (INFN); Monbusho
International Scientific Research Program, JSPS and Toray Science Foundation
of Japan; the Dutch Foundation for Fundamenteel Onderzoek der Materie (FOM);
the U.K. Particle Physics and Astronomy Research Council; and the U.S.
Department of Energy and National Science Foundation.
%

\end{document}

%% file: autori_qm.tex

\def\groupalberta{\affiliation{Department of Physics, University of Alberta, Edmonton, Alberta T6G 2J1, Canada}}
\def\groupargonne{\affiliation{Physics Division, Argonne National Laboratory, Argonne, Illinois 60439-4843, USA}}
\def\groupbari{\affiliation{Istituto Nazionale di Fisica Nucleare, Sezione di Bari, 70124 Bari, Italy}}
\def\groupcolorado{\affiliation{Nuclear Physics Laboratory, University of Colorado, Boulder, Colorado 80309-0446, USA}}
\def\groupdesy{\affiliation{DESY, Deutsches Elektronen-Synchrotron, 22603 Hamburg, Germany}}
\def\groupzeuthen{\affiliation{DESY Zeuthen, 15738 Zeuthen, Germany}}
\def\groupdubna{\affiliation{Joint Institute for Nuclear Research, 141980 Dubna, Russia}}
\def\grouperlangen{\affiliation{Physikalisches Institut, Universit\"at Erlangen-N\"urnberg, 91058 Erlangen, Germany}}
\def\groupferrara{\affiliation{Istituto Nazionale di Fisica Nucleare, Sezione di Ferrara and Dipartimento di Fisica, Universit\`a di Ferrara, 44100 Ferrara, Italy}}
\def\groupfrascati{\affiliation{Istituto Nazionale di Fisica Nucleare, Laboratori Nazionali di Frascati, 00044 Frascati, Italy}}
\def\groupfreiburg{\affiliation{Fakult\"at f\"ur Physik, Universit\"at Freiburg, 79104 Freiburg, Germany}}
\def\groupgent{\affiliation{Department of Subatomic and Radiation Physics, University of Gent, 9000 Gent, Belgium}}
\def\groupgiessen{\affiliation{Physikalisches Institut, Universit\"at Gie{\ss}en, 35392 Gie{\ss}en, Germany}}
\def\groupglasgow{\affiliation{Department of Physics and Astronomy, University of Glasgow, Glasgow G12 8QQ, United Kingdom}}
\def\groupillinois{\affiliation{Department of Physics, University of Illinois, Urbana, Illinois 61801-3080, USA}}
\def\groupmit{\affiliation{Laboratory for Nuclear Science, Massachusetts Institute of Technology, Cambridge, Massachusetts 02139, USA}}
\def\groupmichigan{\affiliation{Randall Laboratory of Physics, University of Michigan, Ann Arbor, Michigan 48109-1120, USA }}
\def\groupmoscow{\affiliation{Lebedev Physical Institute, 117924 Moscow, Russia}}
\def\groupmunich{\affiliation{Sektion Physik, Universit\"at M\"unchen, 85748 Garching, Germany}}
\def\groupnikhef{\affiliation{Nationaal Instituut voor Kernfysica en Hoge-Energiefysica (NIKHEF), 1009 DB Amsterdam, The Netherlands}}
\def\groupstpetersburg{\affiliation{Petersburg Nuclear Physics Institute, St. Petersburg, Gatchina, 188350 Russia}}
\def\groupprotvino{\affiliation{Institute for High Energy Physics, Protvino, Moscow region, 142281 Russia}}
\def\groupregensburg{\affiliation{Institut f\"ur Theoretische Physik, Universit\"at Regensburg, 93040 Regensburg, Germany}}
\def\grouprome{\affiliation{Istituto Nazionale di Fisica Nucleare, Sezione Roma 1, Gruppo Sanit\`a and Physics Laboratory, Istituto Superiore di Sanit\`a, 00161 Roma, Italy}}
\def\groupsimonfraser{\affiliation{Department of Physics, Simon Fraser University, Burnaby, British Columbia V5A 1S6, Canada}}
\def\grouptriumf{\affiliation{TRIUMF, Vancouver, British Columbia V6T 2A3, Canada}}
\def\grouptokyo{\affiliation{Department of Physics, Tokyo Institute of Technology, Tokyo 152, Japan}}
\def\groupamsterdam{\affiliation{Department of Physics and Astronomy, Vrije Universiteit, 1081 HV Amsterdam, The Netherlands}}
\def\groupwarsaw{\affiliation{Andrzej Soltan Institute for Nuclear Studies, 00-689 Warsaw, Poland}}
\def\groupyerevan{\affiliation{Yerevan Physics Institute, 375036 Yerevan, Armenia}}


\groupalberta
\groupargonne
\groupbari
\groupcolorado
\groupdesy
\groupzeuthen
\groupdubna
\grouperlangen
\groupferrara
\groupfrascati
\groupfreiburg
\groupgent
\groupgiessen
\groupglasgow
\groupillinois
\groupmit
\groupmichigan
\groupmoscow
\groupmunich
\groupnikhef
\groupstpetersburg
\groupprotvino
\groupregensburg
\grouprome
\groupsimonfraser
\grouptriumf
\grouptokyo
\groupamsterdam
\groupwarsaw
\groupyerevan


\author{A.~Airapetian}  \groupyerevan
\author{N.~Akopov}  \groupyerevan
\author{Z.~Akopov}  \groupyerevan
\author{M.~Amarian}  \groupzeuthen \groupyerevan
\author{V.V.~Ammosov}  \groupprotvino
\author{A.~Andrus}  \groupillinois
\author{E.C.~Aschenauer}  \groupzeuthen
\author{W.~Augustyniak}  \groupwarsaw
\author{R.~Avakian}  \groupyerevan
\author{A.~Avetissian}  \groupyerevan
% ...

\collaboration{The HERMES Collaboration} \noaffiliation
